\documentclass[twocolumn,showpacs]{revtex4} %
\usepackage{epsfig,amsmath,amssymb}
\usepackage{graphicx}
\usepackage[english]{babel}

\begin{document}

\title{Critical temperature of the superfluid transition  \\
in Fermi-system at an arbitrary pair potential }

\author{Yu.\,M. Poluektov}
\email{yuripoluektov@kipt.kharkov.ua} %
\affiliation{National Science Center ``Kharkov Institute of Physics
and Technology'', Akhiezer Institute for Theoretical Physics, 61108
Kharkov, Ukraine}
\author{A.\,A. Soroka} %
\affiliation{National Science Center ``Kharkov Institute of Physics
and Technology'', Akhiezer Institute for Theoretical Physics, 61108
Kharkov, Ukraine}

\begin{abstract}
A method for calculation of the critical temperature of transition
of a many-particle Fermi system into a superfluid or a
superconducting state at an arbitrary pair potential of the
interparticle interaction is proposed. An original homogeneous
integral equation, that determines the critical temperature, is
transformed into a homogeneous integral equation with a symmetric
kernel that enables application of a general theory of integral
equations for calculation. Examples are given of calculation of the
superconducting transition critical temperature for the BCS
potential with the Coulomb repulsion and the critical temperature of
the superfluid transition of liquid helium-3 into  {\it p}\,-wave
pairing state for the Morse potential. %
\end{abstract}
\maketitle

\section{INTRODUCTION}
\vspace{-4mm}

Calculation of the critical temperature of a superfluid or
superconducting transition in a Fermi system is a central issue in
the theory of superfluid Fermi systems. The importance of this
problem has especially risen after the discovery of high-temperature
superconductivity \cite{HTSC,HTSC2}. In calculation of the critical temperature %
in the theory of conventional superconductivity \cite{BCS,BCS2,BTS},
it is assumed that the interaction between the electrons exists only
in a narrow energy range of the order of the Debye energy near the
Fermi surface. This allows one to approximate the interelectron
interaction potential by a constant. But such approximation becomes
fairly rough to account for the influence of the Coulomb interaction
\cite{BTS,P1,P2}, because in this case the region of interaction is
not confined by a narrow energy range. The validity of
approximations, employed in calculation of the critical temperature
in conventional superconductors, is broken in many cases in
high-temperature superconductors. Of special interest is calculation
of the transition critical temperature in superconductors with
overlapping conduction bands \cite{P1,P2,P3}, because such a model
can be used for a theoretical description of superconductivity both
in the transition metals and high-temperature superconductors. A
detailed analysis of calculations of the superconducting transition
critical temperature in different systems is given in the book
\cite{HTSC3}. It is noted therein (p.\,27) that the solving of an
integral equation determining the critical temperature for in the
least bit complex interaction ``constitutes a well-known problem''.

Taking into account of interparticle interaction only in a narrow
range near the Fermi energy in calculation of the temperature of
transition of $^3$He into the superfluid phase \cite{3He1} is also,
generally speaking, inconsistent. In particular, apparently this is
responsible for a difference of some orders of magnitude in
theoretical estimations of the $^3$He critical temperature in papers
of different authors \cite{3He2,3He3}. Besides that, the reason of
such large discrepancies is connected with sensitivity of the
calculated temperature to a choice of parameters of the
interparticle interaction potential.

The aim of this paper is to develop a consistent method for
calculation of the critical temperature of transition of a
many-particle Fermi system into a superfluid or a superconducting
state at an arbitrary pair potential of the interparticle
interaction. The essence of the proposed method consists in
transition to a study of a homogeneous integral equation with a
symmetric kernel that enables application of the known results of
the theory of integral equations. In principle, the proposed method
enables to calculate numerically the critical temperature for an
arbitrary interparticle interaction potential with a prescribed
accuracy. As examples of use of the proposed method, the critical
temperature in the BCS model with a finite ratio of the Debye energy
to the Fermi energy is calculated and the influence of the Coulomb
repulsion is taken into account. It is shown that the formula for
the critical temperature \cite{BTS,SC} in the presence of the
Coulomb interaction, obtained earlier by means of an approximate
consideration, is exact within the framework of the selected model.
The critical temperature of the superfluid transition of liquid
$^3$He into {\it p}\,-wave pairing state for the Morse potential is
calculated and its strong dependence on the value of one of
parameters of this potential is shown. At reasonable choice of
parameters of the potential, the calculated critical temperature is
close to the experimentally observable one.

\vspace{-2mm}
\section{TRANSFORMATION OF THE EQUATION FOR THE CRITICAL TEMPERATURE TO %
A HOMOGENEOUS INTEGRAL EQUATION WITH A SYMMETRIC KERNEL } %
\vspace{-3mm}

The critical temperature $T_C$  of transition into a superfluid or a
superconducting state of a Fermi system is found
from the linearized equation for the order parameter $\Delta_{\bf k}$ \cite{SC}: %
\\ \vspace{-3mm} %
\begin{equation} \label{EQ01}
\begin{array}{c}
\displaystyle{%
\Delta_{\bf k}=-\frac{1}{V}\sum_{\bf k'} U_{{\bf k}-{\bf k'}}\frac{\Delta_{\bf k'}}{2\,\xi_{k'}}\tanh\frac{\xi_{k'}}{2T_C} \,, %
}
\end{array}
\end{equation}
where ${\bf k}$  is the wave vector, $\xi_k=\hbar^2k^2\big/2m_*
-\mu$, $\mu$ is the chemical potential, $m_*$ is the effective mass.
Fourier component of the interaction potential has the form
\begin{equation} \label{EQ02}
\begin{array}{c}
\displaystyle{ %
U_{\bf k}=\int\!\! d{\bf r}\, U\!({\bf r})\,e^{-i {\bf k}{\bf r}}, %
}
\end{array}
\end{equation}
where $U\!({\bf r})$ is a potential energy of the interparticle
interaction. We will consider an isotropic system, and expand the
order parameter and the Fourier component of the potential in
spherical functions
\begin{equation} \label{EQ03}
\begin{array}{c}
\displaystyle{ %
\Delta_{\bf k}=\sqrt{4\pi}\sum_{l=0}^\infty\sum_{m=-l}^l \Delta_{lm}\!(k)\,Y_{lm}\!(\Omega), %
}
\end{array}
\end{equation}
\vspace{-3mm}
\begin{equation} \label{EQ04}
\begin{array}{c}
\displaystyle{ %
U_{{\bf k}-{\bf k'}}=4\pi\sum_{l=0}^\infty\sum_{m=-l}^l U_l(k,k')\,Y_{lm}\!(\Omega)\,Y_{lm}^*\!(\Omega'). %
}
\end{array}
\end{equation}
Supposing that the potential depends only on a distance $U\!({\bf r})=U\!(r)$, %
we find for a component of the potential
\begin{equation} \label{EQ05}
\begin{array}{c}
\displaystyle{ %
U_l(k,k')=4\pi\int_0^\infty\! d{r}\,r^2 U\!(r) j_l(kr) j_l(k'r), %
}
\end{array}
\end{equation}
$\displaystyle{j_l(x)=\sqrt{\frac{\pi}{2x}}J_{l+\frac{1}{2}}(x) }$ %
are Bessel spherical functions. In the case of the singlet pairing,
when $\Delta_{\bf k}\!=\!\Delta_{-\bf k}$, in the expansion
(\ref{EQ03}) only components with even $l$ remain, and for the
triplet pairing, when $\Delta_{\bf k}\!=\!-\Delta_{-\bf k}$, only
components with odd $l$. In an anisotropic medium, for example in a
crystal, the expansion of the order parameter and the potential
should be made in basis functions of the irreducible representations
of a symmetry group of a~crystal~\cite{P4}.\nolinebreak[4]

Equation (\ref{EQ01}) decomposes into equations for the separate
components of expansions (\ref{EQ03}),(\ref{EQ04}) with different
$l$ for the order parameter and the potential, which after the
transition in (\ref{EQ01}) from summation to integration acquire the
form
\begin{equation} \label{EQ06}
\begin{array}{c}
\displaystyle{ %
\Delta_l(k)=-\frac{1}{(2\pi)^2} \int_0^\infty\! U_l(k,k')F(k';T_C)\Delta_l(k')k'^2\,dk', %
}
\end{array}
\end{equation}
where $\displaystyle{\Delta_l(k)\equiv\sum_{m=-l}^l a_{lm}\Delta_{lm}(k),\,a_{lm} }$ %
are arbitrary coefficients, because the transition temperature does
not depend on the index $m$. In (\ref{EQ06}) the designation is
introduced
\begin{equation} \label{EQ07}
\begin{array}{c}
\displaystyle{ %
F(k)\equiv F(k;T_C)\equiv \frac{1}{\xi_k}\tanh\frac{\xi_k}{2T_C} \,. %
}
\end{array}
\end{equation}
Thus, calculation of the critical temperature comes to finding
values of the parameter $T_C$, at which the homogeneous integral
equation
\begin{equation} \label{EQ08}
\begin{array}{c}
\displaystyle{ %
\Delta(k)=\int_0^\infty\! R(k,k')\Delta(k')\,dk' %
}
\end{array}
\end{equation}
has different from zero solutions. The kernel of this integral
equation
\begin{equation} \label{EQ09}
\begin{array}{c}
\displaystyle{ %
R(k,k')= -\frac{1}{(2\pi)^2}\,U\!(k,k')F(k';T_C)k'^2 %
}
\end{array}
\end{equation}
is, obviously, unsymmetric $R_l(k,k')\neq R_l(k',k)$. The index $l$
of the order parameter and the potential here and in what follows is
omitted.

For subsequent analysis it is convenient to go over to an integral
equation with a symmetric kernel, that will allow us to use general
results of the theory of integral equations. For this purpose we
multiply equation (\ref{EQ06}) by $k\sqrt{F(k)}$ and define the function %
$\Phi(k)\equiv k\sqrt{F(k)}\Delta(k)$. %
As a result we come to the homogeneous integral equation
\begin{equation} \label{EQ10}
\begin{array}{c}
\displaystyle{ %
\Phi(k)=\int_0^\infty\! K(k,k';T_C)\Phi(k')\,dk' , %
}
\end{array}
\end{equation}
with the symmetric real kernel
\begin{equation} \label{EQ11}
\begin{array}{c}
\displaystyle{ %
K(k,k';T_C)\equiv K(k,k')=K(k',k)= %
}\vspace{2mm}\\%
\displaystyle{ %
= -\frac{kk'}{(2\pi)^2}\,U\!(k,k')\sqrt{F(k)F(k')} \,.%
}%
\end{array}
\end{equation}
As opposed to the Fredholm homogeneous integral equation of the
second kind \cite{Krasnov1}
\begin{equation} \label{EQ12}
\begin{array}{c}
\displaystyle{ %
\varphi(x)=\lambda\int_a^b\! K(x,t)\varphi(t)\,dt , %
}
\end{array}
\end{equation}
in equation (\ref{EQ10}) a parameter, which role is played by the
critical temperature here, enters not as a factor before the
integral, but is incorporated in a complicated manner into the
kernel of the integral equation $K(k,k';T_C)$.

Note that the order parameter $\Delta(k)$ is, generally speaking,
complex, so that it and the above defined function can be presented in the form %
$\Delta(k)=\Delta'(k)+i\Delta''(k)$ and $\Phi(k)=\Phi'(k)+i\Phi''(k)$. %
Because of the kernel (\ref{EQ11}) being real, identical integral
equations come out for the real and the imaginary parts of the function $\Phi(k)$. %
Therefore, in what follows we consider this function real.

Despite the fact that equation (\ref{EQ10}) is not of the Fredholm
form, finding of the critical temperature can be reduced to finding
of characteristic numbers $\lambda$ of the equation of type
(\ref{EQ12}). To this end, let's consider the auxiliary equation
\begin{equation} \label{EQ13}
\begin{array}{c}
\displaystyle{ %
\Phi(k)=\lambda(T)\int_0^\infty\! K(k,k';T)\Phi(k')\,dk' . %
}
\end{array}
\end{equation}
The kernel of this equation depends on the parameter $T$, which in
our case has the meaning of temperature. Characteristic numbers of
equation (\ref{EQ13}) $\lambda_i(T)$ (several numbers can exist) are
functions of temperature. Equation (\ref{EQ13}) coincide with
equation (\ref{EQ10}) in the case when at some temperature one of
its characteristic numbers appears to be equal to unity, so that the
critical temperature has to satisfy the condition
\begin{equation} \label{EQ14}
\begin{array}{c}
\displaystyle{ %
\lambda_i(T_C)=1. %
}
\end{array}
\end{equation}
From here on assuming eigenfunctions to be normed by the condition %
$\displaystyle{\int_0^\infty\! \Phi^2(k)\,dk =1}$, we get %
\begin{equation} \label{EQ15}
\begin{array}{c}
\displaystyle{ %
\lambda(T)= \bigg[\int_0^\infty\! \Phi(k)K(k,k';T)\Phi(k')\,dk'dk\bigg]^{-1} . %
}
\end{array}
\end{equation}
Transition to a superfluid state is possible only for such
potentials at which equation (\ref{EQ13}) has positive
characteristic numbers. From equation (\ref{EQ15}) and the form of
function (\ref{EQ07}), it follows that characteristic numbers are
large at high temperatures and fall off with decreasing temperature.
A phase transition takes place when condition (\ref{EQ14}) becomes
satisfied for one of numbers. If for a given potential condition
(\ref{EQ14}) is not satisfied at any temperature, then a phase
transition is not possible. Dependence of the characteristic number
on temperature for the potential of the BCS type (\ref{EQ20}) is
shown in figure~1. Qualitatively, dependence of absolute values of
characteristic numbers on temperature has a similar form as well for
arbitrary potentials.
\vspace{-3mm} %
\begin{figure}[h!]
\centering %
\includegraphics[width = 1.12\columnwidth]{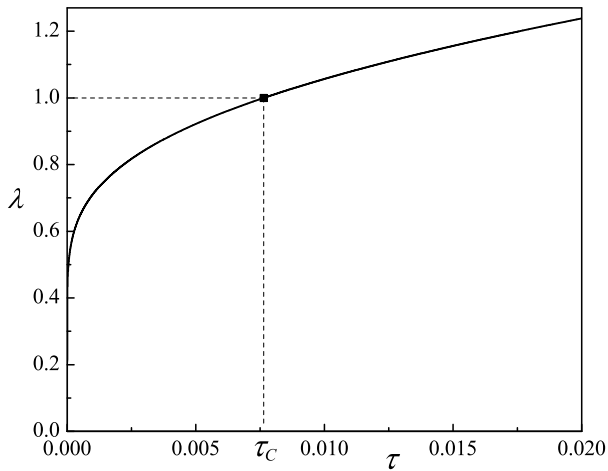} %
\vspace{-7mm}
\caption{\label{fig01} %
Dependence of the characteristic number on temperature in the BCS model, %
$\displaystyle{\lambda(\tau)=\bigg( \frac{g}{2} \int_{-1/\tau}^{1/\tau}\! dy\,\frac{1}{y}\tanh\frac{y}{2} \bigg)^{-1}}$, %
$\displaystyle{\tau\equiv\frac{T}{\varepsilon_D},\, g=0.2}$. %
The mark $\scriptstyle{\blacksquare}$ designates the point,
corresponding to transition to a superfluid state: %
$\lambda(\tau_C=1)$.
}%
\end{figure}

\section{CRITICAL TEMPERATURE FOR SEPARABLE POTENTIALS} %

Characteristic numbers of the integral equation (\ref{EQ13}) can
easily be found if, for instance, its kernel can be represented as a
product of the two same factors
\begin{equation} \label{EQ16}
\begin{array}{c}
\displaystyle{ %
K(k,k')= \nu\,K(k)K(k'), %
}
\end{array}
\end{equation}
where $\nu=\pm 1$. Such a representation takes place for potentials
of the special form $U(k,k')=-\nu\,U(k)U(k')$ , called separable
potentials. In this case
\begin{equation} \label{EQ17}
\begin{array}{c}
\displaystyle{ %
K(k)=\frac{k}{2\pi}\,U(k)\sqrt{F(k)} %
}%
\end{array}
\end{equation}
and the integral equation has the single characteristic number
\begin{equation} \label{EQ18}
\begin{array}{c}
\displaystyle{ %
\lambda(T)= \nu\bigg[\int_0^\infty\! K^2(k)\,dk\bigg]^{-1} . %
}
\end{array}
\end{equation}
This number is positive if $\nu=+1$. In the opposite case the
characteristic number is negative and a phase transition is absent.
Condition (\ref{EQ14}) leads to the equation determining the
critical temperature for the case of the separable potentials
\begin{equation} \label{EQ19}
\begin{array}{c}
\displaystyle{ %
\frac{1}{(2\pi)^2}\int_0^\infty\! U^2(k)F(k;T_C)k^2\,dk=1. %
}%
\end{array}
\end{equation}
Note that the separable model potentials are widely used, for
instance, in the theory of the nucleus \cite{Eisenberg}.

The matrix element of the interaction potential in the BCS model can
be represented as
\begin{equation} \label{EQ20}
\begin{array}{c}
\displaystyle{ %
U(k,k')= -U_0\psi(k)\psi(k'), %
}
\end{array}
\end{equation}
where $\psi(k)=\theta(k-k_F-\Delta k)-\theta(k-k_F+\Delta k)$, and
$k_F=(3\pi^2n)^{1/3}$ is the Fermi wave number, $n$ is the electron
density, $\theta(k)$ is the unit step function, $U_0>0$. The width
of the shell nearby the Fermi surface, in which fermions interact,
is determined by the Debye frequency $\omega_D$, so that %
$\Delta k=m_*\omega_D\big/\hbar k_F$. As seen, the BCS potential is
separable, with $\nu=1$ and $U(k)=\sqrt{U_0}\,\psi(k)$, and
therefore the integral equation (\ref{EQ13}) has the single
characteristic number. In this case equation (\ref{EQ19}) can be
represented in a standard for the BCS theory \cite{BCS,BCS2,BTS}
form
\begin{equation} \label{EQ21}
\begin{array}{c}
\displaystyle{ %
\frac{U_0}{(2\pi)^2}\int\limits_{k_F-\Delta k}^{k_F+\Delta k}\frac{k^2}{\xi_k}\tanh\!\frac{\xi_k}{2T_C}\,dk =1. %
}%
\end{array}
\end{equation}
In calculation of the transition temperature into a superconducting
state one needs, generally, account for the two energy parameters
characterizing a system, namely the Debye energy $\varepsilon_D=\hbar\omega_D$ %
and the Fermi energy $\varepsilon_F=\hbar^2k_F^2\big/2m_*$. %
In conventional superconductors the Fermi energy is by two orders
higher than  the Debye energy, so that $\varepsilon_F\gg \varepsilon_D$. %
In some cases, for instance in high-temperature superconductors
\cite{HTSC,HTSC2}, this strong inequality can be violated. In this
connection, let's make calculation of the critical temperature, not
assuming the ratio $\Gamma\equiv\varepsilon_D\big/\varepsilon_F$ to
be small, so that equation (\ref{EQ21}) can be written in the form
\begin{equation} \label{EQ22}
\begin{array}{c}
\displaystyle{ %
\frac{U_0N_F}{2}\int\limits_{-\varepsilon_{\!D}\!\big/\!T_C}^{\varepsilon_{\!D}\!\big/\!T_C}
   \!\!\sqrt{1+\frac{T_C}{\varepsilon_F}\,y}\,\,\frac{1}{y}\tanh\!\frac{y}{2}\,\,dy =1, %
}%
\end{array}
\end{equation}
$N_F=m_*k_F\big/2\pi^2\hbar^2$ is the density of states at the Fermi
surface. Usually, assuming that $\Gamma\ll 1$, the square root in
formula (\ref{EQ22}) is assumed to be unity. In addition, in further
calculation in the BCS theory it is assumed that $\varepsilon_D\gg T_C$. %
This condition is also violated in the HTSC. A standard BCS formula
for the superconducting transition temperature, calculated under the
condition $\varepsilon_D\gg T_C$, has the form \cite{SC}
\begin{equation} \label{EQ23}
\begin{array}{c}
\displaystyle{ %
T_C=\frac{2\gamma}{\pi}\,\varepsilon_D\exp\!\!\bigg[ -\frac{1}{U_0N_F} \bigg], %
}%
\end{array}
\end{equation}
where $\gamma=e^C$, $C\approx 0.577$ is the Euler's constant. In
general case, the critical temperature is determined by zero of the
function
\begin{equation} \label{EQ24}
\begin{array}{c}
\displaystyle{ %
\Psi_p\!\left(\frac{T}{\varepsilon_D},\Gamma,g\right)\equiv
1-\frac{g}{2}\int\limits_{-\varepsilon_{\!D}\!\big/\!T}^{\varepsilon_{\!D}\!\big/\!T}
   \!\!\sqrt{1+\frac{T}{\varepsilon_D}\,\Gamma\,y}\,\,\frac{1}{y}\tanh\!\frac{y}{2}\,\,dy \,, %
}%
\end{array}
\end{equation}
depending on the three dimensionless parameters $T\big/\varepsilon_D$, %
$\Gamma\equiv \varepsilon_D\big/\varepsilon_F$ and $g=U_0N_F$. %
Dependence of the critical temperature on the parameter $\Gamma$ %
at a fixed value of $g$ is shown in figure 2. At maximum possible
value $\Gamma=1$, the critical temperature decreases by nearly $10\%$. %
\vspace{-4mm}
\begin{figure}[h!]
\centering %
\includegraphics[width = 1.05\columnwidth]{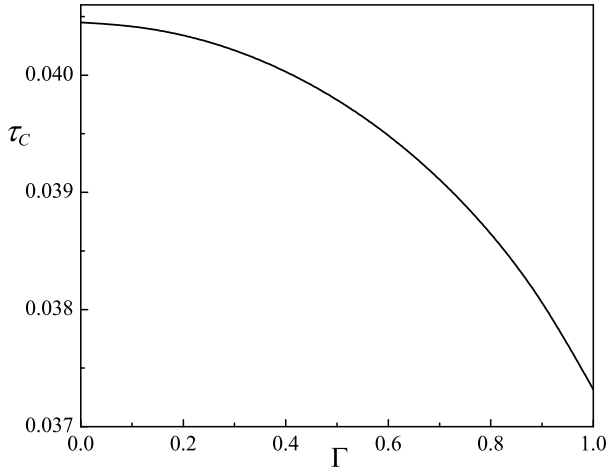} %
\vspace{-7mm}
\caption{\label{fig02} %
Dependence of the critical temperature on the ratio of the Debye
energy to the Fermi energy: %
$\displaystyle{\tau_C\equiv T_C\big/\varepsilon_D =\tau_C(\Gamma),\, g=0.3}$. %
}%
\end{figure}

\section{CRITICAL TEMPERATURE IN THE CASE OF THE DEGENERATE KERNEL} %

If the potential can be represented in the form
\begin{equation} \label{EQ25}
\begin{array}{c}
\displaystyle{ %
U(k,k')= -\sum_{i=1}^N \nu_i\,U_i(k)U_i(k'), %
}
\end{array}
\end{equation}
where $\nu_i=+1$ or $\nu_i=-1$ and $N$ is a finite number, then the
kernel in the integral equation (\ref{EQ13}) is
\begin{equation} \label{EQ26}
\begin{array}{c}
\displaystyle{  %
K(k,k')= \sum_{i=1}^N \nu_i\,K_i(k)K_i(k'), %
}
\end{array}
\end{equation}
and besides
\begin{equation} \label{EQ27}
\begin{array}{c}
\displaystyle{ %
K_i(k)= U_i(k)\,\frac{k}{2\pi}\sqrt{F(k)}\,. %
}%
\end{array}
\end{equation}
For such kernel (\ref{EQ26}), which is called degenerate, finding of
the characteristic numbers $\lambda$ or the corresponding
eigenvalues $\sigma\equiv 1/\lambda$ is reduced to finding of roots
of the algebraic equation \cite{Krasnov2}
\begin{equation} \label{EQ28}
\begin{array}{c}
\displaystyle{ \Delta(\sigma)\equiv \left|
                   \begin{array}{cccc}
                      \nu_1K_{11}-\sigma & \nu_2K_{12}        & \dots & \nu_NK_{1N}        \vspace{4mm}\\ %
                      \nu_1K_{21}        & \nu_2K_{22}-\sigma & \dots & \nu_NK_{2N}        \vspace{4mm}\\ %
                      \dots              & \dots              & \dots & \dots              \vspace{4mm}\\%
                      \nu_1K_{N1}        & \nu_2K_{N2}        & \dots & \nu_NK_{NN}-\sigma \vspace{0mm}%
                   \end{array}\!
                 \right| =0,
}
\end{array}
\end{equation}
where
\begin{equation} \label{EQ29}
\begin{array}{c}
\displaystyle{ %
K_{ij}=K_{ji}= \frac{1}{(2\pi)^2}\int_0^\infty\! U_i(k)U_j(k)F(k)k^2 dk\,. %
}%
\end{array}
\end{equation}
It is obviously that $K_{ii}>0$. As noted above, a phase transition
occurs when, at decreasing temperature, one of the roots becomes
unity in the transition point. If for a given potential that does
not take place, then no phase transition exists.

Let's consider in more detail the simpler case $N=2$, when the
potential is represented in the form
\begin{equation} \label{EQ30}
\begin{array}{c}
\displaystyle{ %
U(k,k')= -\nu_1\,U_1(k)U_1(k')-\nu_2\,U_2(k)U_2(k') \,. %
}
\end{array}
\end{equation}
Then, according to (\ref{EQ28}), the characteristic numbers are
determined from the quadratic equation
\begin{equation} \label{EQ31}
\begin{array}{c}
\displaystyle{ %
\sigma^2-\sigma\big(\nu_1K_{11}+\nu_2K_{22}\big)+\nu_1\nu_2\big(K_{11}K_{22}-K_{12}^2\big)=0, %
}
\end{array}
\end{equation}
having the solutions
\begin{equation} \label{EQ32}
\begin{array}{c}
\displaystyle{ %
\sigma_\pm = \frac{1}{2}\big[ (\nu_1K_{11}+\nu_2K_{22}) \pm D \big] \,, %
}
\end{array}
\end{equation}
where $D=\sqrt{(\nu_1K_{11}-\nu_2K_{22})^2 +4\nu_1\nu_2K_{12}^2 \vphantom{\big)}}$. %
For solutions (\ref{EQ32}), there are obvious relations
\begin{equation} \label{EQ33}
\begin{array}{c}
\displaystyle{ %
\sigma_+ -\sigma_- = D>0, %
}\vspace{2mm}\\%
\displaystyle{ %
\sigma_+ +\sigma_- = \nu_1K_{11}+\nu_2K_{22}, %
}\vspace{2mm}\\%
\displaystyle{ %
\sigma_+^2 -\sigma_-^2 = (\nu_1K_{11}+\nu_2K_{22})D, %
}\vspace{2mm}\\%
\displaystyle{ %
\sigma_+\sigma_- = \nu_1\nu_2(K_{11}K_{22}-K_{12}^2), %
}%
\end{array}
\end{equation}
so that $\sigma_+ > \sigma_-$. Hence, if $\nu_1K_{11}+\nu_2K_{22}<0$ %
and $\nu_1\nu_2\big(K_{11}K_{22}-K_{12}^2\big)>0$, then both
eigenvalues are negative and a phase transition into a superfluid
state is not possible. Whereas if $\nu_1K_{11}+\nu_2K_{22}>0$, %
then for $\nu_1\nu_2\big(K_{11}K_{22}-K_{12}^2\big)<0$ %
one eigenvalue is positive $\sigma_+ > 0$ and another is negative
$\sigma_- < 0$, and for $\nu_1\nu_2\big(K_{11}K_{22}-K_{12}^2\big)>0$ %
both eigenvalues are positive. In both latter cases the critical
temperature is determined by the condition %
$\lambda_+(T_C)=1\big/\sigma_+(T_C)=1$, so that the equation for
$T_C$ is the following
\begin{equation} \label{EQ34}
\begin{array}{lr}
\displaystyle{ \hspace{0mm}%
\nu_1K_{11}\!(T_C)+\nu_2K_{22}(T_C) \, +   %
}\vspace{2mm}\\%
\displaystyle{ \hspace{+4mm}%
+ \sqrt{\big[\nu_1K_{11}\!(T_C)-\nu_2K_{22}(T_C)\big]^2 + 4\nu_1\nu_2K_{12}^2(T_C) }=2.  %
}%
\end{array}
\end{equation}
This equation can be written as well in the form
\begin{equation} \label{EQ35}
\begin{array}{lr}
\displaystyle{ \hspace{-0mm}%
\big[\nu_1K_{11}\!(T_C)+\nu_2K_{22}(T_C)\big] \, -   %
}\vspace{2mm}\\%
\displaystyle{ \hspace{19mm}%
- \nu_1\nu_2\big[K_{11}\!(T_C)K_{22}(T_C)-K_{12}^2(T_C)\big] = 1.  %
}%
\end{array}
\end{equation}
As a concrete example, let's consider calculation of the critical
temperature when, besides the attraction described by the BCS model,
there is also repulsion in a certain shell around the Fermi surface.
Then $\nu_1=1$, $\nu_2=-1$ and $U_1(k)=\sqrt{U_0}\,\psi_1(k)$, %
$U_2(k)=\sqrt{U_C}\,\psi_2(k)$, and besides %
$\psi_1(k)=\theta(k-k_F-\Delta k_1)-\theta(k-k_F+\Delta k_1)$, %
$\psi_2(k)=\theta(k-k_F-\Delta k_2)-\theta(k-k_F+\Delta k_2)$, %
$\Delta k_1=m_*\omega_D\big/\hbar k_F$, %
$\Delta k_2=m_*\omega_C\big/\hbar k_F$. %
Here $\omega_C\approx\varepsilon_F\big/\hbar$, so that %
$\Delta k_2 > \Delta k_1$. This model approximately describes the
influence of the Coulomb repulsion on the superconducting transition
temperature. From equation (\ref{EQ35}) it follows
\begin{equation} \label{EQ36}
\begin{array}{c}
\displaystyle{ %
K_{11}\!(T_C)= 1+\frac{K_{12}^2(T_C)}{1+K_{22}(T_C)}\,, %
}%
\end{array}
\end{equation}
where in the present case
\begin{equation} \label{EQ37}
\begin{array}{lrc}
\displaystyle{\hspace{0mm} %
K_{11}\!(T_C)= \sqrt{\frac{U_0}{U_C}}\,K_{12}(T_C)= %
}\vspace{1mm}\\
\displaystyle{\hspace{14mm} %
=\frac{U_0N_F}{2}\int\limits_{-\varepsilon_{\!D}\!\big/\!T_C}^{\varepsilon_{\!D}\!\big/\!T_C}
   \!\!\sqrt{1+\frac{T_C}{\varepsilon_F}\,y}\,\,\frac{1}{y}\tanh\!\frac{y}{2}\,\,dy\,,  %
}\vspace{3mm}\\
\displaystyle{ %
K_{22}(T_C)=
\frac{U_CN_F}{2}\int\limits_{-\varepsilon_{\!F}\!\big/\!T_C}^{\varepsilon_{\!F}\!\big/\!T_C}
   \!\!\sqrt{1+\frac{T_C}{\varepsilon_F}\,y}\,\,\frac{1}{y}\tanh\!\frac{y}{2}\,\,dy \,. %
}%
\end{array}
\end{equation}
As shown above, even for maximum $\Gamma\equiv\varepsilon_D\big/\varepsilon_F =1$ %
at replacement of the root under integrals (\ref{EQ37}) by unity the
difference from the exact result is no more than $10\%$. From this
approximation it follows
\begin{equation} \label{EQ38}
\begin{array}{c}
\displaystyle{ %
K_{11}= g_0\Theta,\,\,\, K_{22}= g_C\!\left(\!\Theta+\ln\!\frac{\varepsilon_F}{\varepsilon_D} \right),\,\,\, %
K_{12}= \sqrt{g_0g_C}\,\Theta ,
}%
\end{array}
\end{equation}
where $g_0\equiv U_0N_F$, $g_C\equiv U_CN_F$,
$\displaystyle{\Theta\equiv\ln\!\frac{2\gamma\varepsilon_D}{\pi T_C}}$.  %
Then from (\ref{EQ36}) the equation for the critical temperature follows  %
\\ \vspace{-4mm} %
\begin{equation} \label{EQ39}
\begin{array}{c}
\displaystyle{%
\Theta \bigg[ g_0 - \frac{g_C}{1+g_C\ln(\varepsilon_F/\varepsilon_D) } \bigg] =1,  %
}%
\end{array}
\end{equation}
which coincides exactly with the one obtained in the model under
discussion earlier \cite{BTS,SC}. Note, that in \cite{SC} (p.~129)
it is stressed that the integral equation determining the critical
temperature is very difficult to solve even for a simple
interaction, and therefore the consideration carried out in
\cite{SC} proves to be ``quite approximate''.  However, as shown
above, formula (\ref{EQ39}) is exact for the selected model
interaction.

\section{CRITICAL TEMPERATURE FOR PAIR POTENTIALS OF GENERAL FORM} %

If we have such a potential that the kernel of integral equation
(\ref{EQ13}) cannot be represented in the form of (\ref{EQ16}) or
(\ref{EQ26}), then there exists an infinite number of characteristic
numbers. In this case we have to employ approximate methods for
finding characteristic numbers of the symmetric kernels
\cite{Krasnov2,KK,MS}. Reasonably efficient is the Ritz method, and
we will employ it for finding eigenvalues (characteristic numbers)
of the nondegenerate kernels. Since in the
problem under study eigenfunctions are defined on the half-axis %
$0\leq k < \infty$, it is convenient to expand them in the functions %
$l_n(k/k_F\!)\equiv\exp(-k/2k_F\!)L_n(k/k_F\!)$, which are expressed
through the Laguerre polynomials $L_n(k/k_F\!)$ \cite{NU}:
\begin{equation} \label{EQ40}
\begin{array}{c}
\displaystyle{ %
\Phi(k)=\sum_{n=0}^\infty B_n\,l_n(k/k_F\!) \,. %
}
\end{array}
\end{equation}
Since the orthogonality relation holds for the introduced functions, %
$\int_0^\infty l_n(k)l_m(k)dk = \delta_{nm}$ , then the coefficients
of the expansion (\ref{EQ40}) are defined by formula
\begin{equation} \label{EQ41}
\begin{array}{c}
\displaystyle{ %
B_n = \frac{1}{k_F}\int_0^\infty\! \Phi(k)\,l_n(k/k_F\!)\,dk \,. %
}
\end{array}
\end{equation}
From the normalization condition it follows $\sum_{n=0}^\infty\! B_n^2\!=\!1$. %
Expansion (\ref{EQ40}) holds for functions satisfying the
requirement of square integrability \cite{NU}:
\begin{equation} \label{EQ42}
\begin{array}{c}
\displaystyle{ %
\int_0^\infty\!\Phi^2(k)\,e^{-k/k_F}dk < \infty \,. %
}
\end{array}
\end{equation}
Then, the integral equation (\ref{EQ13}) is reduced to the system of
an infinite number of homogeneous linear equations
\begin{equation} \label{EQ43}
\begin{array}{c}
\displaystyle{ %
\sum_{n=0}^\infty\!\big[\sigma\delta_{mn} - K_{mn} \big] B_n =0, %
}
\end{array}
\end{equation}
where
\begin{equation} \label{EQ44}
\begin{array}{lr}
\displaystyle{\hspace{0mm} %
K_{nm}=K_{mn}=-\frac{1}{(2\pi)^2}\int\limits_0^\infty\!\!\int\limits_0^\infty U(k,k')\,e^{-\frac{k+k'}{2k_F}}\times %
}\vspace{2mm}\\%
\displaystyle{\hspace{11mm} %
\times\,L_n\!\!\left(\!\frac{k}{k_F}\!\right)\!L_m\!\!\left(\!\frac{k'}{k_F}\!\right)\!\!\sqrt{F(k)F(k')}\,kk'dkdk' .%
}%
\end{array}
\end{equation}
The eigenvalues are determined by the condition of the determinant
equaling zero
\begin{equation} \label{EQ45}
\begin{array}{c}
\displaystyle{ %
\textrm{det}\big[\sigma\delta_{mn} - K_{mn} \big] =0\,. %
}
\end{array}
\end{equation}
In the Ritz method the infinite determinant is replaced by the
determinant of the finite matrix. Since a phase transition takes
place under satisfaction of the condition %
$\sigma(T_C)=\lambda^{-1}(T_C)=1$, then the equation for the
critical temperature of a phase transition takes the form
\begin{equation} \label{EQ46}
\begin{array}{c}
\displaystyle{ %
\textrm{det}\big[\delta_{mn} - K_{mn}(T_C) \big] =0\,. %
}
\end{array}
\end{equation}

If in expansion (\ref{EQ40}) only two main terms with $n=0$ and
$n=1$ are left, then from (\ref{EQ45}) it follows the quadratic
equation for approximate determination of the two highest by
absolute value eigenvalues
\begin{equation} \label{EQ47}
\begin{array}{c}
\displaystyle{ %
\sigma^2-\sigma(K_{00}+K_{11}\!)+K_{00}K_{11}-K_{01}^2=0. %
}
\end{array}
\end{equation}
Hence
\begin{equation} \label{EQ48}
\begin{array}{c}
\displaystyle{ %
\sigma_\pm = \frac{1}{2}\big( K_{00}+K_{11} \pm D \big) , %
}
\end{array}
\end{equation}
where $D=\sqrt{(K_{00}-K_{11})^2 +4K_{01}^2 \vphantom{\big)}}$\,. %
From obvious relations for solutions (\ref{EQ48})
\begin{equation} \label{EQ49}
\begin{array}{c}
\displaystyle{ %
\sigma_+ -\sigma_- = D>0,\quad \sigma_+ +\sigma_- = K_{00}+K_{11}, %
}\vspace{2mm}\\%
\displaystyle{ %
\sigma_+^2 -\sigma_-^2 = (K_{00}+K_{11}\!)D,\quad  \sigma_+\sigma_- = K_{00}K_{11}-K_{01}^2 %
}
\end{array}
\end{equation}
it follows that, if $K_{00}+K_{11}<0$  and $K_{00}K_{11}-K_{01}^2>0$, %
then both roots (\ref{EQ48}) are negative and a phase transition is absent. %
Whereas if $K_{00}+K_{11}>0$, then one (for $K_{00}K_{11}-K_{01}^2<0$) %
or two (for $K_{00}K_{11}-K_{01}^2>0$) positive roots exist, and the
transition temperature is determined from the condition $\sigma_+(T_C)=1$, %
so that
\begin{equation} \label{EQ50}
\begin{array}{lr}
\displaystyle{ \hspace{0mm}%
K_{00}(T_C)+K_{11}(T_C) \, +   %
}\vspace{2mm}\\%
\displaystyle{ \hspace{+6mm}%
+\,\sqrt{\big[K_{00}(T_C)-K_{11}(T_C)\big]^2 + 4K_{01}^2(T_C) }=2,  %
}%
\end{array}
\end{equation}
or
\begin{equation} \label{EQ51}
\begin{array}{lr}
\displaystyle{\hspace{0mm}%
\big[K_{00}(T_C\!)+K_{11}\!(T_C\!)\big] \!-\! \big[K_{00}(T_C\!)K_{11}\!(T_C\!)\!-\!K_{01}^2\!(T_C\!)\big] = 1.  %
}%
\end{array}
\end{equation}

Consider one more method of approximate determination of the
critical temperature not requiring use of the eigenfunctions. For a
symmetric kernel under the normalization condition %
$\int_0^\infty\!\Phi_i(k)\Phi_j(k)\,dk = \delta_{ij}$ the bilinear
expansion holds \cite{Krasnov1}:
\begin{equation} \label{EQ52}
\begin{array}{c}
\displaystyle{ %
K(k,k')=\sum_{j=1}^N \frac{\Phi_j(k)\,\Phi_j(k')}{\lambda_j} . %
}
\end{array}
\end{equation}
Here the number of characteristic numbers $N$ can be finite for
degenerate kernels and infinite in general case. It is assumed that
the following conditions are satisfied for the kernel
\begin{equation} \label{EQ53}
\begin{array}{c}
\displaystyle{ %
\int\limits_0^\infty\!\! K(k,k';T_C)dkdk'<\infty,\,\, %
\int\limits_0^\infty\!\! K^2(k,k';T_C)dkdk'<\infty, %
}
\end{array}
\end{equation}
i.e. the integral equation is of Fredholm type. If kernel
(\ref{EQ52}) is continuous and its characteristic numbers are all
positive, or there is only a finite number of negative
characteristic numbers, then the bilinear series of this kernel
converges uniformly (Mercer's theorem) \cite{Krasnov1,Krasnov2}.
Assuming in (\ref{EQ52}) $k\!=\!k'$ and integrating both sides by
$k$, we obtain the formula for the trace of the continuous kernel
$K(k,k')$:
\begin{equation} \label{EQ54}
\begin{array}{c}
\displaystyle{ %
\int_0^\infty\!\! K(k,k)dk\, = \sum_{j=1}^N \frac{1}{\lambda_j} \,. %
}%
\end{array}
\end{equation}
Squaring both sides of equation (\ref{EQ52}) and integrating by $k$
and $k'$, we come as well to the relation
\begin{equation} \label{EQ55}
\begin{array}{c}
\displaystyle{ %
\int\limits_0^\infty\!\!\int\limits_0^\infty\! K^2(k,k')dkdk' = \sum_{j=1}^N \frac{1}{\lambda_j^2} \,. %
}%
\end{array}
\end{equation}
Taking into account that $\lambda_1(T_C)=1$ in the transition point,
formulae (\ref{EQ54}) and (\ref{EQ55}) can be used for both an
approximate calculation of the critical temperature and checking
self-consistency of calculations of the critical temperature by
other approximate methods. If the maximum eigenvalue $\sigma_1=1/\lambda_1$ %
is positive and $\sigma_1\gg |\sigma_2|>|\sigma_3|>\dots$, %
then the equation
\begin{equation} \label{EQ56}
\begin{array}{c}
\displaystyle{ %
\int_0^\infty\!\! K(k,k;T_C)dk\, = 1, %
}%
\end{array}
\end{equation}
following from (\ref{EQ54}), can be used for calculation of $T_C$.
However, this approximation is by no means always valid. It can turn
out that, in addition to the maximum positive eigenvalue, a
substantial contribution is made by another eigenvalue which, if
negative, can even exceed by absolute value a contribution of the
positive eigenvalue. Therefore, for obtaining more accurate equation
for $T_C$, at least the two maximum by absolute value eigenvalues
$\sigma_1$ and $\sigma_2$ should be taken into account. From
(\ref{EQ54}) and (\ref{EQ55}), the system of algebraic equations
follows for these eigenvalues
\begin{equation} \label{EQ57}
\begin{array}{c}
\displaystyle{\hspace{-0.5mm} %
\sigma_1\!+\sigma_2\!=\!K_1\!\equiv\!\int\limits_0^\infty\!\! K(k,k)dk = \!-\frac{1}{(2\pi)^2}\!\int\limits_0^\infty\!\! U(k,k)F(k)k^2 dk, %
}%
\end{array}
\end{equation}
\begin{equation} \label{EQ58}
\begin{array}{lr}
\displaystyle{\hspace{-2mm} %
\sigma_1^2+\sigma_2^2=\!K_2\equiv\int\limits_0^\infty\!\!\int\limits_0^\infty\! K^2(k,k')\,dkdk' = %
}\vspace{1mm}\\%
\displaystyle{\hspace{8mm} %
= \frac{1}{(2\pi)^4}\int\limits_0^\infty\!\!\int\limits_0^\infty\! U^2(k,k')F(k)F(k')k^2k'^2 dkdk'. %
}%
\end{array}
\end{equation}
The solutions of this system of equations are
\begin{equation} \label{EQ59}
\begin{array}{c}
\displaystyle{ %
\sigma_{1,2} = \frac{1}{2}\big( K_1 \pm D \big) \,, %
}
\end{array}
\end{equation}
where $D=\sqrt{2K_2-K_1^2 \vphantom{ \big) }}$\,. %
Here the subradical expression $2K_2-K_1^2>0$, because all
characteristic numbers of an integral equation with a symmetric
kernel are real. The obvious relations hold
\begin{equation} \label{EQ60}
\begin{array}{c}
\displaystyle{ %
\sigma_1 -\sigma_2 = D,\quad\! \sigma_1^2 - \sigma_2^2 = K_1D, \quad\!  %
\sigma_1\sigma_2 = \frac{1}{2}\big( K_1^2-K_2 \big).
}%
\end{array}
\end{equation}
If $K_1<-D$ and $K_1^2-K_2>0$, then both roots are negative and a
phase transition is absent.  If $K_1>-D$ and $K_1^2-K_2<0$, %
then $\sigma_1>0$ and $\sigma_2<0$, and besides, if in this case
$K_1<0$, then $|\sigma_2|>\sigma_1$. If $K_1>-D$ and $K_1^2-K_2>0$, %
then both roots are positive and $\sigma_1>\sigma_2$. Thus, with
account of the two characteristic numbers the equation for the
critical temperature has the form
\begin{equation} \label{EQ61}
\begin{array}{c}
\displaystyle{%
K_1(T_C) + \sqrt{ 2K_2(T_C)-K_1^2(T_C) } =2 \,, %
}%
\end{array}
\end{equation}
or, equivalently,
\begin{equation} \label{EQ62}
\begin{array}{c}
\displaystyle{%
K_1(T_C) = 1 \pm \sqrt{ K_2(T_C)-1 \vphantom{\big)} } \,. %
}%
\end{array}
\end{equation}
In the latter formula the plus sign should be taken if $K_1^2-K_2>0$, %
and the minus sign if $K_1^2-K_2<0$.

For estimation of value of the minimum (by absolute value)
characteristic number $\lambda_i(T_C)$ $(i\neq 1)$, the method of
iterated kernels \cite{Krasnov1,Krasnov2,MS} can be used, for
instance. The iterated kernel is defined by the relation
\begin{equation} \label{EQ63}
\begin{array}{lr}
\displaystyle{\hspace{0mm}%
K_m(k,k')\equiv \int_0^\infty\! K(k,k_1)K(k_1,k_2)\dots %
}\vspace{2mm}\\%
\displaystyle{\hspace{24mm} %
\dots K(k_{m-1},k')\,dk_1dk_2\dots dk_{m-1},  %
}
\end{array}
\end{equation}
and its trace by the relation
\begin{equation} \label{EQ64}
\begin{array}{lr}
\displaystyle{\hspace{0mm}%
S_m\equiv \int_0^\infty\! K_m(k,k)\,dk = \sum_{i=1}^N \frac{1}{\lambda_i^m} \,. %
}%
\end{array}
\end{equation}
In the transition point
\begin{equation} \label{EQ65}
\begin{array}{lr}
\displaystyle{\hspace{0mm}%
S_m(T_C) \equiv \int_0^\infty\! K_m(k,k;T_C)\,dk = 1+\sum_{i=2}^N \frac{1}{\lambda_i^m(T_C)} \,. %
}%
\end{array}
\end{equation}
Formula (\ref{EQ65}) enables to estimate the value of the minimum
different from unity characteristic number $|\lambda_2(T_C)|$. %
If $|\lambda_2(T_C)|>1$, then
\begin{equation} \label{EQ66}
\begin{array}{lr}
\displaystyle{\hspace{0mm}%
|\lambda_2(T_C)| = \big[|S_m(T_C)-1|\big]^{-1/m} . %
}%
\end{array}
\end{equation}
And if $|\lambda_2(T_C)|<1$, then
\begin{equation} \label{EQ67}
\begin{array}{lr}
\displaystyle{\hspace{0mm}%
|\lambda_2(T_C)| = \big[|S_m(T_C)|\big]^{-1/m} . %
}%
\end{array}
\end{equation}
Formulae (\ref{EQ66}), (\ref{EQ67}) can be used for checking
self-consistency of calculation of the critical temperature by
approximate methods.

\section{TEMPERATURE OF TRANSITION INTO HELIUM-3 SUPERFLUID PHASE FOR THE MORSE POTENTIAL} %

Now we apply the proposed calculation methods to evaluate the
critical temperature of the superfluid transition of liquid $^3$He
into {\it p}\,-wave $(l=1)$ pairing state, using the known Morse
potential
\begin{equation} \label{EQ68}
\begin{array}{lr}
\displaystyle{%
U(r)= u_0\Big\{\!\exp\!\!\big[\!-\!2(r-r_0)/L\big] - 2\exp\!\!\big[\!-\!(r-r_0)/L\big]\!\Big\}, %
}%
\end{array}
\end{equation}
which is widely used for modeling of interactions between atoms and
has the three parameters $u_0,r_0,L$. The parameters $r_0$ and $u_0$
define the minimum point of the potential (a distance between atoms
and a depth of the potential well in this point) and the parameter
$L$ characterizes a scale of potential variation. Fourier component
of potential (\ref{EQ68}) has the form
\begin{equation} \label{EQ69}
\begin{array}{lr}
\displaystyle{%
U_1(k,k')= %
}\vspace{2mm}\\%
\displaystyle{\hspace{2mm}%
=\!\frac{\pi u_0L^3}{2}\Bigg[\!
\exp\!\!\left[\!\frac{2r_0}{L}\!\right]\!I\!\!\left[\!\frac{kL}{2},\frac{k'L}{2}\!\right] %
\!-\!16\exp\!\left[\frac{r_0}{L}\right]\!I\!\left[ kL,k'L \right]\!\Bigg], %
}
\end{array}
\end{equation}
where
\begin{equation} \label{EQ70}
\begin{array}{lr}
\displaystyle{\hspace{0mm}%
I(x,y)\equiv %
}\vspace{2mm}\\%
\displaystyle{\hspace{4mm}%
\equiv\!\frac{\big[1+x^2+y^2\big]}{xy\big[1\!+\!(x\!-\!y)^2\big]\!\big[1\!+\!(x\!+\!y)^2\big]} %
-\frac{1}{4x^2y^2}\ln\!\frac{1\!+\!(x\!+\!y)^2}{1\!+\!(x\!-\!y)^2}.
}%
\end{array}
\end{equation}
In this case the kernel
\begin{equation} \label{EQ71}
\begin{array}{lr}
\displaystyle{\hspace{-2mm}%
K(k,k')= -\frac{u_0L^3}{8\pi}\Bigg[\!
\exp\!\!\left[\!\frac{2r_0}{L}\!\right]\!I\!\!\left[\!\frac{kL}{2},\frac{k'L}{2}\!\right]- %
}\vspace{2mm}\\%
\displaystyle{\hspace{12mm}%
-\,16\exp\!\left[\frac{r_0}{L}\right]\!I\!\left[ kL,k'L \right]\!\Bigg]kk'\sqrt{F(k)F(k')} %
}
\end{array}
\end{equation}
is not degenerate and approximate methods should be used for
calculation of the critical temperature.

For numerical calculation of the $^3$He superfluid transition the
following values of the potential parameters are used: %
$r_0=2.95\!\times\!10^{-8}\,\textrm{cm}$, $u_0/k_\textrm{B}=10.0\,\textrm{K}$. %
The parameter $L$ is chosen from such a criterion that the
calculated scattering length
\begin{equation} \label{EQ72}
\begin{array}{lr}
\displaystyle{\hspace{0mm}%
a_0= \frac{m}{4\pi\hbar^2}\int\! U\!(r)\,d{\bf r}=\frac{u_0m}{4\hbar^2}L^3e^{r_0\!/L}\!\left(e^{r_0\!/L}\!-\!16\right)  %
}%
\end{array}
\end{equation}
should be close to the average adopted in the literature value
$a_0=-8.2\,\textrm{{\AA}}$ \cite{JDK}. It gives for the
dimensionless
parameter $\tilde{L}\equiv L\big/r_0$ the value $\tilde{L}=0.386$. %
The following parameters of the liquid helium are chosen: the
density of atoms $n=1.8\!\times\!10^{22}\,\textrm{cm}^{-3}$, %
that corresponds to the mass density
$\rho_m=0.08\,\,\textrm{g}\!\cdot\!\textrm{cm}^{-3}$, %
the Fermi wave number and the Fermi temperature
$k_F\!=\!0.81\!\times\!10^{8}\,\textrm{cm}^{-1}$, $T_F\!=\!1.71\,\textrm{K}$. %
Numerical calculation by the Ritz method in the approximation of two
characteristic numbers give the value $\,\,\,\tau_C=6.604\!\times\!10^{-4}$, 
that corresponds to the critical temperature $T_C\approx 1.1\,\textrm{mK}$. %
Note that experimentally observable transition temperature increases
from the value $T_C\approx 0.9\,\textrm{mK}$ at zero pressure to the
value $T_C\approx 2.6\,\textrm{mK}$ at the maximum possible pressure
\cite{3He2}. Calculation by formula (\ref{EQ48}) of the second
eigenvalue gives the value $\sigma_-(\tau_C)=-12.8$. In order to
check self-consistency of the result obtained above by the Ritz
method, let's compare the value of the characteristic number
$\lambda_-(\tau_C)=\sigma_-(\tau_C)^{-1}=-0.078$ %
with the value of this number calculated independently by the trace
method (\ref{EQ67}). Result of calculation by the trace method
converges rather rapidly with increasing a number of iterations $m$
and for $m=3$ gives the value $\lambda_-^{(3)}(\tau_C)=-0.082$. %
This result differs from that one obtained above by the Ritz method
by nearly $5\%$, that indicates good accuracy of this approximation
with regard to the Morse potential.

Let's calculate the critical temperature of the phase transition by
another method based on using the relations (\ref{EQ54}),
(\ref{EQ55}) with account of two eigenvalues, that leads to
equations (\ref{EQ61}), (\ref{EQ62}). For the same set of parameters
of the system and the potential as that used
above, we get $\tau_C=1.43\!\times\!10^{-2}$ and $T_C\approx 24.5\,\textrm{mK}$. %
The second eigenvalue, obtained by formula (\ref{EQ59}),
$\sigma_2(\tau_C)\!=\!-16.4$ and the corresponding characteristic
number $\lambda_2(\tau_C)=\sigma_2(\tau_C)^{-1}\!\!=\!-0.061$. %
\!Calculation by the trace method gives
$\lambda_2^{(3)}\!(\tau_C)\!=\!-0.077$ for $m=3$. In this case the
characteristic numbers obtained by different ways differ by nearly
$20\%$. \!Therefore, the second method gives a less accurate result
than the Ritz method and leads to the overestimated value of the
superfluid transition temperature.

It should be noted that the value of the critical temperature proves to %
be very sensitive to that of the dimensionless parameter
$\tilde{L}$. As shown in Table\,\,I, variation of this parameter by
only $0.001\,(\approx\!0.25\%$) leads to variation of $T_C$ by       
two-three times. Such strong dependence can be connected with the
fact that the parameter $\tilde{L}$ of the Morse potential defines
the relation between quantities of the attractive and repulsive
parts of the potential. \!\!With increasing $\tilde{L}$ the role of
the repulsive part of the potential falls and of the attractive part
rises exponentially, and this presumably causes high sensitivity of
the transition temperature to this parameter.

\newpage
\begin{table}[h!] \nonumber
\centering %
\caption{Dependence of the critical temperature and scattering
length on the parameter $\tilde{L}\!=\!L\big/r_0$ of the Morse potential} %
\vspace{0.5mm}%
\begin{tabular}{|c|c|c|c|} \hline
$ \rule{7mm}{0pt} \tilde{L}$ \rule{7mm}{0pt}           & \rule{3mm}{0pt} 0.385 \rule{3mm}{0pt} & \rule{3mm}{0pt} 0.386 \rule{3mm}{0pt} & \rule{3mm}{0pt} 0.387 \rule{3mm}{0pt} \\ \hline %
$T_C\,\,\,\textrm{mK}$    &  0.35        &  1.1        &  2.3      \\ \hline %
$a_0\,\,\,\textrm{{\AA}}$ &  $-7.9$      &  $-8.2$     &  $-8.5$   \\ \hline %
\end{tabular}  
\end{table}

\section{CONCLUSIONS} %

A method for calculation of the temperature of a phase transition of
a many-particle Fermi system into a superfluid or a superconducting
state, based on transition to a homogeneous integral equation with a
symmetric kernel, is proposed. For potentials bringing to a
degenerate kernel of an integral equation the problem of
determination of the critical temperature is solvable exactly, and
for potentials of general form approximate methods of finding
eigenvalues of homogeneous integral equations with symmetric kernels
can be employed. The developed method enables to calculate the
temperature of transition into the superfluid state for an arbitrary
pair potential with a desired accuracy. As  examples of use of the
method, the results of the BCS model with account of a finite ratio
of the Debye energy to the Fermi energy are reproduced and  the
influence of the Coulomb interaction on the critical temperature is
analyzed. It is shown that the known formula \cite{BTS,SC} for the
critical temperature in the BCS model accounting for the Coulomb
interaction is exact for the selected model. The temperature of the
superfluid transition of liquid $^3$He into {\it p}\,-wave pairing
state for the Morse potential is calculated. It is shown that the
calculated temperature depends strongly on the value of one of
parameters of the potential and at reasonable choice of parameters
calculation gives the value of the critical temperature close to the
experimentally observable one. The developed method of calculation
can be employed for determination of temperatures of phase
transitions in other physical systems.


\newpage
\begin{thebibliography}{99}

\bibitem{HTSC}
J.G. Bednorz, A.K. Muller, Possible high   superconductivity in the
Ba--La--Cu--O system, Z. Phys. B. \textbf{64}\,(2), 189--193 (1986).

\bibitem{HTSC2}
L.P. Gor'kov, N.B. Kopnin, High-Tc superconductors from the
experimental point of view, Usp. Fiz. Nauk. \textbf{156}\,(1),
117--135 (1988).

\bibitem{BCS}
J. Bardeen, L. Cooper, J. Schrieffer, Microscopic theory of
superconductivity, Phys. Rev. \textbf{106}\,(1), 162--164 (1957).

\bibitem{BCS2}
J. Bardeen, L. Cooper, J. Schrieffer, Theory of superconductivity,
Phys. Rev. \textbf{108}\,(5), 1175--1204 (1957).

\bibitem{BTS}
N.N. Bogoliubov, V.V. Tolmachev, and D.V. Shirkov, A new method in
the theory of superconductivity, Izd. AN SSSR, Moscow, 128 p.
(1958).

\bibitem{P1}
Yu.M. Poluektov, Two-band model of high- superconductor, Fiz. Nizk.
Temp \textbf{15}\,(4), 362--\,367 (1989).

\bibitem{P2}
Yu.M. Poluektov, Critical temperature in the two-band model of a
superconductor with account of the Coulomb repulsion, %
Preprint: CNII-Atom-Inform. KIPT., Moscow, 88\,-20, 8 p. (1988).

\bibitem{P3}
Yu.M. Poluektov, Critical temperature of two-band superconductor
taking into account interband pairing, Fiz. Nizk. Temp.
\textbf{18}\,(7), 683--\,687 (1992).

\bibitem{HTSC3}
V.L. Ginzburg, D.A. Kirzhnits (Editors), Problem of high-temperature
superconductivity: collection of articles, Nauka, Moscow, 400 p. (1977). %

\bibitem{3He1}
D.D. Osheroff, W.J. Gully, R.C. Richardson, D.M. Lee, New magnetic
phenomena in liquid 3Не below 3 mK, Phys. Rev. Lett.
\textbf{29}\,(14), 920--\,923 (1972).

\bibitem{3He2}
I.M. Khalatnikov (Editor), Superfluidity of helium-3: %
\,\,collection of articles, Mir, Moscow, 288 p. (1977). 

\bibitem{3He3}
A.J. Leggett, A theoretical description of the new phases of liquid
3He, Rev. Mod. Phys. \textbf{47}\,(2), 331--\,414 (1975).

\bibitem{SC}
P.-G. de Gennes, Superconductivity of metals and alloys, Mir,
Moscow, 280 p. (1968).

\bibitem{P4}
Yu.M. Poluektov, To the phenomenological theory of two-band crystal
superconductors with tetragonal and orthorhombic symmetries, Fiz.
Nizk. Temp. \textbf{19}\,(3), 256--267 (1993).

\bibitem{Krasnov1}
M.L. Krasnov, Integral equations, Nauka, Moscow, 304 p. (1975).

\bibitem{Eisenberg}
J. Eisenberg, W. Greiner, Microscopic theory of the nucleus, Mir,
Moscow, 488 p. (1976).

\bibitem{Krasnov2}
M.L. Krasnov, A.I. Kiselev, G.I. Makarenko, Integral equations.
Problems and exercises, Nauka, Moscow, 216 p. (1976).

\bibitem{KK}
L.V. Kantorovich, V.I. Krylov, Approximate methods of higher
analysis, Izd. fiz.-mat. lit., Moskow-Leningrad, 708 p. (1962).

\bibitem{MS}
S.G. Mikhlin, K.L. Smolitskiy, Approximate methods for solution of
differential and integral equations, Nauka, Moskow, 384 p. (1965).

\bibitem{NU}
A.F. Nikiforov, V.B. Uvarov, Foundations of the theory of special
functions, Nauka, Moskow, 304 p. (1974).


\bibitem{JDK}
M.J. Jamieson, A. Dalgarno, M. Kimura, Scattering lengths and
effective ranges for He-He and spin-polarized H-H and D-D scattering
Phys. Rev. A. \textbf{51}, 2626--2629 (1995).

\end{thebibliography}
\end{document}